\documentstyle[12pt,amsmath,amsfonts,amssymb,graphicx]{report}

\begin{document}
\begin{flushright}
IC/00/115
\end{flushright}
\vspace{2cm}
\def\simlt{\stackrel{<}{{}_\sim}}
\def\simgt{\stackrel{>}{{}_\sim}}
\begin{center}
{\Large \bf The Neutron EDM in the SM : A Review} 
\end{center}
\vspace{1cm}
\begin{center}
{\bf  Shahida Dar}
\end{center}
\begin{center}
{\bf The Abdus Salam International Center for Theoretical Physics,I-34100 Trieste, Italy}
\end{center} 
\begin{abstract}
We review the status of the electric dipole moment (EDM) of neutron in the Standard Model (SM).
The contributions of the strong and electroweak interactions are discussed seperately. In each case
the structure of the Lagrangian and the sources of CP violation are specified, and calculational
details are given subsequently. These two contributions to the neutron EDM exist in any extension
of the SM including supersymmetry, two--doublet models as well as models with more than three
generations of fermions. We do not discuss the EDM in such extensions; however, we briefly
summarize their predictions with a detailed account of the related literature.
\end{abstract}
\baselineskip = 0.6cm
\tableofcontents\nonumber


\chapter*{Introduction}
\addcontentsline{toc}{chapter}{Introduction}
The electric dipole moment (EDM), $\vec d$, of a classical charge distribution $\rho(\vec x)$ is given by 
\begin{equation}
\vec d=\int d^3 x \vec{x}\rho(\vec x)~.
\end{equation}
which is a polar vector. It is known that the elementary particles have no intrinsic vector quantity other
than their spin. Therefore, their electric ($\vec d$) as well as magnetic dipole ($\vec \mu$) 
moments are expected to be proportional to their spins ($\vec{S}$). So under discrete spacetime 
transformations, these moments and the spin must have the same transformation properties.

Despite their similarity concerning the particle spin, the electric and magnetic moments have quite different characteristics. 
This can be understood by studying the particle under electromagnetic field. 
Depicted in Fig. 1 is a particle (grey blob) with spin $\vec S$ under electric ($\vec E$) and
magnetic ($\vec B$) fields. Consider the upper part of the figure. Initially, particle is in a frame 
where $\vec{B} || \vec{S}$. Under under time--reversal (T) operation (where $\vec{B}\rightarrow -\ \vec{B}$
and  $\vec{S}\rightarrow -\ \vec{S}$) they still remain parallel. Therefore, the corresponding moment 
of interaction, that is, the magnetic moment ($\vec{\mu}$) implies no violation of the discrete symmetry 
T$\equiv$ CP, assuming CPT is a good symmetry of the Nature. 

Consider now the lower part of the same figure where initial configuration is similar to that of 
the upper part; $\vec{E} || \vec{S}$. However, under a T operation $\vec{E}\rightarrow \vec{E}$
and  $\vec{S}\rightarrow -\ \vec{S}$, and thus, the corresponding interaction term in the Lagrangian 
violates the CP symmetry. Therefore, unlike the magnetic moment discussed above, the electric dipole 
moment implies the violation of CP invariance in Nature. It has long been known that CP is not a 
conserved symmetry of the Nature. Indeed, the CP--odd long--living neutral kaon $K_L$ is known 
to disassociate into the CP--even final state consisting of two neutral pions $\pi^{0} \pi^{0}$ \cite{cr,crnew}.
Therefore, CP is not a respected symmetry at all so all particles with spin, must have an EDM 
at some level.\\

\begin{figure}[ht]
\begin{center}
\includegraphics[angle=0, width=7cm]{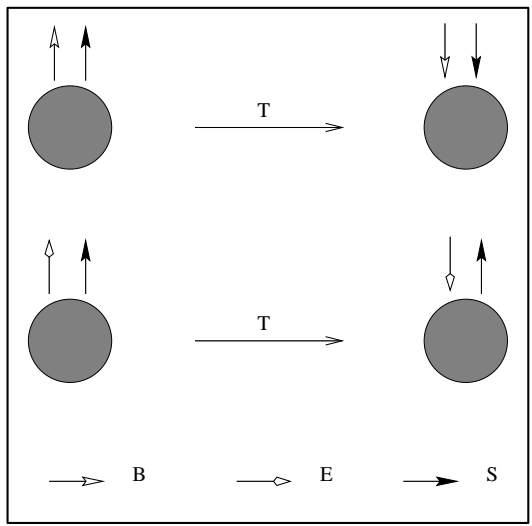}
\caption{\sf Illustration of $\vec{S}\cdot \vec{B}$ (up) and  $\vec{S}\cdot \vec{E}$ (down) interactions 
under time--reversal (T $\equiv$ CP). The former refers to the $magnetic$ $dipole$ $moment$ of the particle whereas
latter to the $electric$ $dipole$ $moment$. Clearly a finite EDM implies the existence of
the CP--violating operators in the underlying theory.}
\end{center}
\end{figure}

At the field theory level, the P and T violating interaction between a spin 1/2 fermion field $\psi(x)$ and the 
photon $A_{\mu}$ can be written as:
\begin{equation}
-\ D(k^2)\overline \psi \left(q+k/2\right)\sigma_{\mu\nu}\gamma_5 k^{\nu} \psi\left(q-k/2\right)\ A^{\mu}(k)~
\end{equation}
where $k$ is the momentum taken away by the photon. One notes that when $\gamma_5$ is removed from this operator the corresponding
quantity becomes the  magnetic moment interaction which does not imply the discrete symmetries. The value of the form factor $D(k^{2})$
for on--shell photon ($k^{2}=0$) is the sought expression for the EDM of the fermion $\psi$:
\begin{equation}
d_\psi\equiv - D(k^{2}=0)
\end{equation}
Of all the baryons the EDM of the neutron \footnote{The neutron is most convenient object for 
precision measurements of the EDMs, owing to vanishing of its electric charge, its sufficient 
stability, and the possibility of using beams of ultracold particles.} $d_n$ has been of interest to the physicists 
for a long time. In 1950, Purcell and Ramsey \cite{pu} first considered the problem of existence of the neutron EDM. 
However, at that time, it was assumed that the physical world is invariant under P inversion, and in order to have a 
nonzero $d_n$, Purcell and Ramsey had to construct a rather unconventional P even EDM of neutron. They gave the 
first limit on the EDM of the neutron:
\begin{equation}
d_n=|\vec d_n|\leq 10^{-13}\mbox{--}3.10^{-18}~{\rm e-cm}~.
\end{equation}
In 1957, Landau \cite{ld} observed that a non-vanishing EDM of the neutron was a signal of P and T violation. 
About that time it was discovered that P invariance is not a strict law of nature \cite{tl} as was experimentally 
confirmed \cite{expwu}. EDM of neutron  violates both parity and time reversal invariance.\\
\hspace*{1.00cm}Even after the parity was found to be violated by the weak interactions it was believed that EDM of the neutron 
was time--reversal invariant and by virtue of CPT theorem was also invariant under CP symmetry. In 1964, when the 
CP violation was observed in kaon system \cite{cr}, the subject of EDM of neutron has become of particular importance 
for both theoretical and experimental interests.\\
\hspace*{1.00cm}Various methods have been invoked for the experimental investigation of the EDMs of 
elementary particles (for details see, for example, \cite{sp}). So far only the upper limits on the EDM of different 
particles have been obtained \cite{upbound}.\\
\hspace*{1.00cm}The detection of a non-zero EDM for neutron would have extremely important implications. It would be 
the first observation of CP violation outside the kaon system. So, it could give us a new understanding of the CP
violation mechanism and of the early universe cosmology where it is generally believed that CP violation gave rise to 
the observed baryon--antibaryon asymmetry of the universe \cite{wi*,goran}. \\
\hspace*{1.00cm}Since the experimental discovery of CP noninvariance in Nature, many diverse mechanisms of the CP 
violation have been proposed and several estimates of the EDM of neutron have been obtained. A comprehensive review 
of the EDM of neutron in various models of CP violation can be found in \cite{pakvasa}.\\
\hspace*{1.00cm}In this dissertation we shall consider the CP violation and hence the neutron EDM in the frame work 
of $SU(3)_c\times SU(2)_L\times U(1)_Y$ theory of electroweak and strong interactions, $i.e.$, {\bf The Standard Model}. In the
standard model there are two sources of CP violation namely:\\
\\
1. The first source of CP violation in the standard model is related to the properties of strong interaction theory 
described by Quantum Chromodynamics (QCD). It has long been realized that, the nontrivial structure of vacuum in QCD 
and the existence of instantons \cite{tt*} in the non-Abelian gauge theory make it necessary to add to the Lagrangian the 
so--called $\theta$-term, which violates P and CP symmetries. This term for $\theta\sim {cal{O}}(1) $ gives an exceedingly 
large value for the neutron EDM which can be made consistent with the experimental bounds \cite{upbound} if $\theta < 10^{-9}$. 
Several methods have been employed to evaluate the effect of CP violation on neutron EDM. The first investigation on this subject was
carried  out by Baluni \cite{bl}. In Chapter I we will discuss sources of CP violation in the strong interactions together with 
a calculation of the neutron EDM.\\
\\
2. The second source of CP violation in the SM appears in the electroweak sector where the mixing matrix 
of quarks $V_{KM}$, appearing in the charged--current interactions, possess a complex phase which generates
CP--violating observables. For three generations of the chiral fermions this matrix can be parametrized by three Euler 
angles and a single phase, $\delta_{CKM}$  \cite{km,cabibbo}. For two generations, for instance, there is no CP violation 
because the quark mixing matrix can always be made real. The  analysis of the EDM of the neutron 
in the electroweak sector is presented in Chapter II,  where we have adopted the calculational 
techniques given by Nanopoulous {\it et al} \cite{nn*}. 

\chapter{The Neutron EDM: Strong Interactions}
In this chapter we will study first the sources of CP violation in strong interactions. 
For this purpose we will work out the nonperturbative and topological properties of 
the QCD vacuum together with its symmetries. Progressively, we will construct the 
representation--independent CP violation Lagrangian which is the mere source of
CP--violating quantities in the hadron spectrum. Finally we will compute the 
EDM of neutron and compare the theoretical estimate with experimental bounds.

\section{Symmetries of the QCD Lagrangian}
The quantum chromodynamics (QCD) is the gauge theory of quarks and  
gluons, which are believed to be constituents of the hadrons. The 
total action density of the theory can be divided into two distinct parts,
\begin {equation}
\label{lagran}
{\cal{L}}_{QCD}={\cal{L}}_{0}+{\cal{L}}_{\theta}~,
\end {equation}
where ${\cal{L}}_{0}$ describes the quarks (color triplets: $\psi^{i}~,~i=1,2,3$)
and gluons (color octets: $A_{\mu}^{a}~,~a=1,\cdots,8$) together with their interactions
\begin {equation}
\label{l0lag}
{\cal{L}}_{0}=-\frac{1}{2 g_s^2} \mbox{Tr}\bigg{[} F_{\mu\nu} F^{\mu\nu}\bigg{]}+i\bigg{\{}\overline{\Psi}_L
(\partial\!\!\!\!\diagup -i A\!\!\!\!\diagup)\Psi_R +(R\to L)\bigg{\}}-\bigg{(}\overline{\Psi}_R {\cal M}_{q} \Psi_L+h.c\bigg{)}~,
\end {equation}
where $g_s$ is the gauge (here color $SU(3)$) coupling constant. In this expression the gluon field strength tensor reads as
$$ F_{\mu\nu}=\partial_{\mu}A_{\nu}-\partial_{\nu}A_{\mu}+i[A_{\mu},A_{\nu}]$$
where $A_{\mu}=A_{\mu}^{a}\Lambda^{a}$, and $3\times3$ anti--hermitian matrices 
$\Lambda^{a}$ ($a=1,\cdots,8$) are the $SU(3)_c$ generators satisfying  
$ 2 \mbox{Tr}\left(\Lambda^{a}\Lambda^{b}\right)= 2 \delta^{a b}$.

The second and third terms in (\ref{l0lag}) refer to the chiral quarks 
$$\Psi_{R,L}=\big{\{}\frac{1}{2}(1\pm\gamma_{5})\psi^{f}\qquad f=1,\cdots, N_f\big{\}}$$
where $N_f$ is the number of flavours, and each $\psi^{f}$ is a color triplet. The quark
mass matrix $M$, which is in general non--hermitian, is an $N_f\times N_f$ matrix in 
the flavour space.

The second piece in (\ref{lagran}) is given by 
\begin{eqnarray}
\label{theta}
{\cal{L}}_{\theta}=-\ \theta \ {\cal{A}}~,\;\;\;\; \mbox{where}~\;\;\;\;\; {\cal{A}}=\frac{1}{16 \pi^{2}}
\mbox{Tr}\bigg{[}\widetilde{F}_{\mu\nu}
F^{\mu\nu}\bigg{]}
\end{eqnarray}
where $0\leq \theta\leq  2 \pi$ is an angle parameter, and the dual field strength tensor is defined by
$\widetilde{F}_{\mu\nu}=\frac{1}{2} \epsilon_{\mu \nu \alpha \beta} F^{\alpha \beta}$. 

For the coherence of the discussions it is convenient to specify the symmetries of the QCD Lagrangian first. 
The main difference between ${\cal{L}}_{0}$ and ${\cal{L}}_{\theta}$ is that the latter 
does not respect the time--reversal (T$\equiv$CP) invariance. This is clear from the fact that (in the temporal gauge: $A_0\equiv 0$)
the first term in ${\cal{L}}_{0}$ corresponds to $\vec{E}_a^{2}-\vec{B}_a^{2}$ whereas ${\cal{L}}_{\theta}$ does to $\vec{E}_a
\cdot\vec{B}_a$,  where $\vec{E}_a $ and $\vec{B}_a$ are the non--Abelian  electric--like and magnetic--like fields. In fact, it is
this CP--violating character of ${\cal{L}}_{\theta}$ which makes it fundamental for calculating the EDMs, which are inherently CP--odd
quantities.  
 
Furthermore, since the gluons are flavour--diagonal, the massless quarks possess a $U_{L}(N_f)\times U_{R}(N_f)$ global 
symmetry in the massless limit; $M\rightarrow 0$. In practice, compared to the dynamical scale of QCD ($\Lambda_{QCD}\sim 300~{MeV}$)
one can take $u$, $d$ and $s$ quarks approximately massless. To this approximation the global symmetry of (\ref{lagran}) becomes
$U_{L}(3)\times U_{R}(3)$. The diagonal group $U_{L+R}(3)\sim SU(3)_{V}\times U(1)_B$ is also respected by the vacuum state since 
($i$) hadrons obey an approximate $SU(3)_{V}$ symmetry ( $i.e.$ the $Eightfold$--$Way$), and ($ii$) the baryon number is conserved.
The remaining symmetries $U_{L-R}(3)\sim SU(3)_{A}\times U(1)_A$ are not manifest in the hadron degeneracies so that 
the dynamics should be such that the QCD vacuum breaks all these axial symmetries. According to the Goldstone theorem, 
from the break--down of $SU(3)_{A}$ we expect eight pseudo scalars which can indeed be identified with the set
$[\pi^+, \pi^0, \pi^-, K^{+}, K^{0}, \overline{K^{0}}, K^-, \eta]$. However, there is no pseudoscalar meson in the
spectrum to be identified with the Goldstone boson of $U(1)_A$.  Hence, in the massless limit, there should be a remnant global
symmetry of the QCD Lagrangian given by the phase rotation $\psi \rightarrow e^{i\alpha\ \gamma_5}\ \psi_L$. However, this symmetry 
is realized neither linearly (by the vacuum invariance) nor nonlinearly (by the existence of a light pseudoscalar). 

As will  be detailed in the following sections, the resolution of the $U(1)_A$ problem and the appearance of the CP--violating
Lagrangian ${\cal{L}}_{\theta}$ are both related to the nontrivial vacuum structure of QCD. Although there will be explicit 
quoting of the relevant work in the analyses below, still an exhaustive review of these problems can be found in \cite{tt*}
together with \cite{thooft}.

\section{QCD Vacuum}

For a clear understanding of the QCD vacuum it is convenient to discuss the topological 
properties of the continuous functions. To study the topological properties of the 
continuous  functions one can divide them into $homotopic$ $classes$; each class contains 
the functions that can be deformed continuously into each other. As a warm--up case 
one can consider the $S^{1}\rightarrow S^{1}$ mapping. Let $A$ be the set of points $\left\{\alpha \right\}$ 
on a unit circle  $S^{1}$, and let $B$ be the set of unimodular complex numbers $\left\{e^{i \beta}\right\}$.
Then the function $f_{\gamma}\left(\alpha \right)=e^{[i(n \alpha + \gamma)]}$ realizes the mapping $\left\{\alpha \right\}\subset
S^{1} \rightarrow \left\{e^{i \beta}\right\}\subset S^{1}$. For a fixed $n$, $f_{\gamma}\left(\alpha \right)$ forms a homotopy 
class as $\gamma$ varies . More precisely, given $f_{\gamma_0}$ and $f_{\gamma_1}$ with $\gamma_0\neq \gamma_1$ then
these two functions can be deformed continuously into each other by the $homotopy$ $F(\alpha, x)=e^{[i(n \alpha + (1-x) \gamma_0 + x
\gamma_1)]}$ where $x\in \left\{0,1\right\}$. One can visualize $f_{\gamma}\left(\alpha \right)$ as mapping $n$ points
of one circle into one point of the other circle in other words one winds around the latter by $n$ times. Every homotopy 
class is characterized by a $winding$ $number$ 
\begin {equation}
\label{wind}
n=\int_{0}^{2 \pi} \frac{d \alpha}{2 \pi} \frac{-i}{f_{\gamma}\left(\alpha \right)}\frac{d f_{\gamma}\left(\alpha \right)}{d \alpha}
\end {equation}
In particular, mapping with $n=1$ is $f_{\gamma}^{(1)}(\alpha)=e^{i\alpha}$, and one can obtain the mapping for any winding number $n$
by using $\left(f_{\gamma}^{(1)}\right)^{n}$. 

Though the $S^1\rightarrow S^1$ mapping above is useful for illustrating the homotopy classes, a physically relevant case occurs only  when the
symmetry group under concern has an $SU(2)$ subgroup. Here the question arises: ``How much it depends on the 
gauge group being $SU(2)$ ?'' The answer is: Firstly, if the gauge group is $U(1)$, it is easy to see that 
every mapping of $S^3$ into $U(1)$ is continuously deformable into the trivial mapping (all the $S^3$ is 
mapped into a single point). Thus {\it for an abelian gauge theory there is no analog of winding number.} 
Secondly, there is a theorem by Raoul Bott \cite{bott} which states that any continuous mapping of $S^3$ into G can be 
continuously deformed into a mapping into $SU(2)$ subgroup of G, where G is a  simple Lie Group. Thus 
everything which will be discovered for $SU(2)$ will be true for an arbitrary simple Lie Group, in particular 
it will be true for $SU(n)$.\\
As will be clear below, the $SU(2)$ group structure is needed to have 
a one--to--one correspondence with the spherical geometry at spatial infinity. Therefore, we now discuss the 
$S^{3}\rightarrow S^{3}$ mapping
\cite{bpst} -- a mapping from the three--dimensional Euclidean space (parametrized by three Euler angles) into the $SU(2)$ space (characterized by
three parameters) . In fact, we now seek for solutions to the classical $SU(2)$ Yang--Mills theory in Euclidean space
($|\vec{r}|^2=x_0^2+\vec{x}\cdot\vec{x}$) with $finite$ $action$
\begin{equation}
\label{lelag}
{\cal{S}}_{E}=\frac{1}{2 g_s^2}\int d^{4}x\ \mbox{Tr}\bigg{[} F_{\mu\nu} F^{\mu\nu}\bigg{]}~,
\end{equation}
that is, the field strength tensor must vanish at infinity,
\begin{equation}
\label{fmunuvan}
F_{\mu\nu}(x) \underset{|\vec{r}| \rightarrow \infty} \longrightarrow 0 ~.
\end{equation}
Normally this condition is equivalent to vanishing of $A_{\mu}(x)$ at infinity. However, a vanishing field
configuration is equivalent to $U\partial_{\mu} U^{-1}$ with $U\in SU(2)$ thanks to the gauge transformation
\begin{equation}
A_{\mu}^{\prime}=UA_{\mu} U^{-1} + U\partial_{\mu} U^{-1}~.
\end{equation}
Therefore, $F_{\mu\nu}(x)$ still vanishes at infinity (\ref{fmunuvan}) for $A_{\mu}$ in the $pure$ $gauge$; 
$$A_{\mu}(x) \underset{|\vec{r}| \to \infty} \longrightarrow U^{-1} \partial_{\mu} U.$$ One notices that 
the points at infinity in the  Euclidean  space are three--spheres, $S^{3}$. Therefore, the gauge transformation
matrix $U\in SU(2)$ realizes a mapping from $S^{3}(\mbox{Euc. space})$ to the $SU(2)(\mbox{group space})$. More
precisely, the manifold of the $SU(2)$ group elements is topologically equivalent to the three--sphere $S^3$. This
can be seen from the fact that $U\in SU(2)$ can be expressed as
\begin{equation}
U=e^{i \vec{\alpha}\cdot \vec{\sigma}}\leadsto u_0 + \vec{u}\cdot \vec{\sigma}~,\ \ \ \mbox{for}\ \ \ |\vec{\alpha}|< < 1.
\end{equation}
Due to unitary character of the gauge transformations, $U^{\dagger} U=U U^{\dagger}=1$ , it is clear that $u_\mu$ spans a sphere
$$ u_0^2+\vec{u}\cdot \vec{u} =1$$
that is the group space of $SU(2)$ is topologically equivalent to the unit sphere, $S^{3}$. Following Coleman \cite{coleman77}, the
previous expression (\ref{wind}) for the winding number can now be written as 
\begin{eqnarray}
\label{wind3}
n&=&\frac{1}{24 \pi^{2}}\int d^{4} x\ \partial_{\mu}
\epsilon_{\mu\nu\rho\lambda}\mbox{Tr}\bigg{[}A_{\nu}A_{\rho}A_{\lambda}\bigg{]}=\int d^{4}x\ {\cal{A}}
\end{eqnarray}
where $A_{\mu}$ here are in the pure gauge discussed above, and the quantity  ${\cal{A}}$ is defined in (\ref{theta}). This 
is the expression for $S^{3}\rightarrow S^{3}$ winding number corresponding to the $SU(2)$ gauge theory. From this very 
expression of the winding number (\ref{wind3}) it can be shown that the action (\ref{lelag}) is indeed finite, ${\cal{S}}_{E}\geq 8 \pi
n^2/g_s^2$, and the corresponding gauge field is either self--dual or antiself--dual, $i.e.$, $F_{\mu\nu}=\pm \widetilde{F}_{\mu\nu}$.

The finite--action solutions of the classical Euclidean  Yang--Mills system (\ref{lelag}) are $instanton$ solutions \cite{inst} 
$$A_{\mu}=\left(\frac{r^2}{r^2+\gamma^2}\right)U\partial_{\mu} U^{-1}$$ 
where $\gamma$ is an arbitrary scale parameter, and the corresponding gauge transformation matrix has the form 
$$U_{1}(\vec{r})=\frac{1}{|\vec{r}|}\left( x_0 + i \vec{x}\cdot \vec{\sigma}\right)~.$$
Thus, the gauge field is of finite extension, and as $|\vec{r}|\rightarrow \infty$ it approaches the pure gauge form, as expected.
For this very solution the action integral has the value $8 \pi^2/ g_s^2$, that is, this solution corresponds to the homotopy class
$n=1$ hence the subscript "1" in $U_{1}(x)$. Clearly the solution for the $n$th homotopy class can be obtained by compounding $U$
by $n$ times: $$U_{n}(\vec r)=(U_{1}(\vec{r}))^{n}.$$

Suppose that there are two vacua belonging to different homotopy classes $n$ and $m$. Then the vacuum to vacuum transition
amplitude reads as 
\begin{equation}
\langle n|e^{-i H t}|m\rangle=\int [d A_{\mu}]_{n-m} e^{i S}
\end{equation}
In imaginary time (or Euclidean space), this becomes
\begin{eqnarray} 
\label{trans}
\langle n|e^{- H t}|m\rangle &=&\int[d A_{\mu}]_{n-m} e^{- S_{E}}\nonumber\\
&=&\int[d A_{\mu}]_{n-m} e^{-\int d^{4}x [L(A_{\mu})]}\sim e^{- 8 \pi (n-m)^{2}/g_s^2}~.
\end{eqnarray}
where $S_{E}= - i S$ is the Euclidean action.\\ 
Therefore the intanton configuration corresponds to tunneling between the vacuum states having different winding numbers. The rough
estimate above, based on the finiteness of the Eucleadean action, shows clearly that the effect here is inherently
$nonperturbative$, that is, the transition rate is enhanced in the limit of large gauge coupling. Besides this, the rate is
enhanced for neighboring vacua; $|n-m|=1$. In the limit of small gauge coupling, the transition amplitude is diminished leaving
system in any of the (equivalent) vacua with inherently $perturbative$ dynamics. In summary we have infinite number of vacua 
each being characterized by its homotopy class or the winding number. Another point of primary importance is the behavior 
of these vacua under gauge transformations. The gauge transformation that changes $A_{\mu}^{n}$ to $A^{n+1}_{\mu}$ works as
\begin{eqnarray}
A^{n+1}_{\mu}&=&U_{n+1}\partial_{\mu} U_{n+1}^{-1}\nonumber\\
&=&U_{1}U_{n}\partial_{\mu} U_{n}^{-1} U_{1}^{-1}\nonumber\\
&=&U_{1}\partial_{\mu} U_{1}^{-1}+U_{1}A_{\mu}^{n}U_{1}^{-1}
\end {eqnarray}
so that the collection of the vacua $\left\{|n\rangle\right\}$ is not gauge--invariant at all:  
\begin {equation}
U_{1}|n \rangle=|n+1 \rangle~.
\end{equation}
Thus, these infinitely many vacua are connected by topologically nontrivial gauge transformations.

Since the vacuum states $|n\rangle$ belonging to different homotopy classes (having different winding numbers) are separated by the 
energy barriers (as dictated by the tunneling amplitude in (\ref{trans})), it is clear that the $true$ vacuum state will be a
linear superposition of all these vacua. Similar to the situation with periodic potentials in quantum mechanics (whose 
true ground state is the Bloch wave), we define the true vacuum to be 
\begin{equation}
|\theta\rangle\equiv\sum_{n} e^{-i n \theta}\ |n\rangle
\end{equation}
which we name as the ``$\theta-vacuum$''. As expected, in contrast to $\left\{|n\rangle\right\}$ the $\theta-vacuum$ is
gauge--invariant. This can be seen by computing 
\begin {eqnarray}
U_{1}|\theta \rangle&=&U_{1}\sum_{n} e^{- i n \theta}|n \rangle \nonumber\\
&=&\sum_{n} e^{- i n \theta}|n+1 \rangle \nonumber\\
&=&\sum_{n} e^{i \theta}e^{- i n \theta}|n+1 \rangle \nonumber\\
&=& e^{i \theta}|\theta \rangle~.
\end {eqnarray}
Therefore, $\theta$ labels the physically inequivalent sectors of the theory, and in each sector we can work physical
processes in a gauge--invariant manner. The different $\theta$ worlds do not  communicate with each other unlike
$\left\{|n\rangle\right\}$ which can be communicated  via the quantum tunneling. Obviously, $\theta$ is arbitrary, and thus, 
there is no $a$ $priori$ way of determining it from the theory. 

The amplitude for a transition between two $\theta$ worlds in the presence of an external conserved current $J$ is given by 
\begin {eqnarray}
\langle\theta^{'}|e^{-i H t}|\theta\rangle_{J}&=&\sum_{m , n}e^{i m \theta}e^{- i n \theta} ~
~\langle m|e^{-i H t}|n\rangle_{J}\nonumber\\                                   
&=&\sum_{m , n} e^{-i(n-m) \theta}e^{i m (\theta^{'}-\theta)}\int[d A_{\mu}]_{n-m} e^{i \int d^{4}x\ (L+J A)}\nonumber\\
&=&\delta(\theta^{'}-\theta)\sum_{\nu\equiv n-m}e^{- i \nu \theta}\int[d A]_{\nu} e^{i \int
d^{4}x\ (L+JA)}\nonumber\\ 
&=&\delta(\theta^{'}-\theta)\sum_{\nu}\int[d A]_{\nu} e^{i\int d^{4}x\ (L_{eff}+JA)}
\end{eqnarray}
where in the last step we have introduced the effective action density
\begin {equation}
L_{eff}=L+\Delta L~\ \ \ \ \mbox{with}\ \ \ \ \Delta L = -\ \theta \nu\equiv -\ g_s^{2}\ {\cal{L}}_{\theta}
\end{equation}
where ${\cal{L}}_{\theta}$ is already defined in (\ref{theta}). Stating in more explicit terms, the entire effect of 
the $\theta$ vacuum can be taken into consideration by introducing the effective action 
\begin {equation}
S_{eff}=-\ \frac{1}{2} \int d^{4}x\ \mbox{Tr}\bigg{[} F_{\mu\nu} F^{\mu\nu}\bigg{]} -\ \theta \frac{\alpha_{s}}{4 \pi} \int d^{4}x
\mbox{Tr}\bigg{[}\widetilde{F}_{\mu\nu} F^{\mu\nu}\bigg{]}
\end{equation} 
where we have rescaled the gauge field as  $A_\mu\rightarrow g_s \ A_\mu$, for convenience.  It is clear that this
effective action is equivalent to (\ref{l0lag}). 

In concluding this section we note that the QCD vacuum has a nontrivial structure described by the gauge field 
configuration corresponding to ``pure gauge''. The finite action solutions of the Euclidean $SU(2)$ Yang--Mills
system (\ref{l0lag}) split into infinite number of equivalent subsets corresponding to the homotopy classes
of $S^{3}(\mbox{Euclidean space})\rightarrow S^{3}(\mbox{SU(2) space})$ mapping. The gauge field configurations
are $instanton$ solutions in that they correspond to tunneling between the vacua with different winding 
numbers. Therefore, the Yang--Mills system is characterized by infinite number of vacua communicated
by quantum tunneling. Moreover, the vacua are not gauge invariant; gauge transformation with winding number $\nu$
transforms the vacuum state $|n\rangle$ to $|n+\nu\rangle$. 

The true ground state of the theory is given by a linear superposition of these vacua whereby defining the
so--called $\theta$ vacuum. The $\theta$ vacuum is gauge--invariant, and each value of the parameter $\theta$
corresponds to a physically distinct state of the system. This parameter is in general arbitrary the theory 
gives no way to determine it.

There is no communication among distinct $\theta$ vacua; moreover, the corresponding transition amplitude
is described by an effective action which differs from the original Yang--Mills action by ${\cal{L}}_{\theta}$
in (\ref{theta}). The most interesting property of this additional piece is that it is odd under the CP
transformation. Therefore, the appearance of  ${\cal{L}}_{\theta}$ is due to the nontrivial nonperturbative 
topological structure of the QCD vacuum.

\section{Effective Vacuum Angle}

As was mentioned when discussing the symmetries of the QCD Lagrangian, there is an excess axial global symmetry ($U(1)_A$) 
of (\ref{l0lag}) which has no imprint in the hadron spectrum. A resolution of the $U(1)_A$ problem rests on the 
anomalous nature of this global symmetry and the nontrivial structure of the QCD vacuum discussed above. The conserved
current, for massless $u$, $d$, and $s$ quarks,  reads as 
\begin {equation}
J_{5}^{\mu}=\overline u \gamma^{\mu}\gamma_{5} u + \overline d \gamma^{\mu}\gamma_{5} d + \overline s \gamma^{\mu}\gamma_{5} s~.
\end {equation}
At tree approximation this current is conserved because its divergence, which is proportional to the quark masses, vanishes identically
in the massless limit. However, this current is no longer conserved if one goes to one loop level
\begin {equation}
\label{abj}
\partial_{\mu} J_{5}^{\mu}=2\ N_{f}\ g_s^2\ {\cal{A}}
\end {equation}  
where $N_f=3$ for the present case. This result is independent of the looping quark masses. Moreover, this result
is exact, that is, it does get no contribution from higher loops (the famous ABJ anomaly \cite{adler,bj}). It is this
quantum mechanical nonconservation of $J_{5}^{\mu}$ which forbids an $exact$ $U(1)_{A}$ symmetry \cite{thooft2}. In this 
sense the QCD Lagrangian has no excess symmetry compared to the observed hadron spectrum, the partonic (quarks and gluons) and
hadronic symmetry properties agree.   

Though this observation solves the $U(1)_A$ problem, there is more to be done with the QCD anomaly of $J_{5}^{\mu}$. First,
one observes that ${\cal{A}}$ can, in fact, be written as a total divergence
\begin{equation}
\partial_{\mu} K^{\mu}=16 \pi^{2}\ {\cal{A}}~\ \ \ \mbox{with}~\ \ \ K_{\mu}= \mbox{Tr}\bigg{[}
\widetilde{F}_{\mu\nu}A_{\nu}\bigg{]}~,
\end{equation}
with, however, the fact that $K^{\mu}$ is $not$ gauge invariant at all. Despite this, one can embed the anomalous divergence in 
(\ref{abj}) to define the a new current
\begin{equation}
\label{new}
\widetilde{J}^{\mu}_{5}=J_{5}^{\mu} - 2\ N_{f}\ \frac{g_s^2}{16 \pi^{2}}\ K^{\mu}
\end{equation}
which is conserved in the massless quark limit. Therefore, the anomalous character of $J_{5}^{\mu}$ (that is the mismatch of the 
classical and quantum symmetries) is now removed via (\ref{new}). However, under a gauge transformation its conserved charge 
\begin {equation}
\widetilde Q_{5}=\int d^{3}x \widetilde J_{5}^{0}
\end {equation}  
can be shown to shift as follows
\begin {eqnarray}
U_{1}\widetilde Q_{5} U_{1}^{-1}&=&U_{1}\int d^{3}x \widetilde J_{5}^{0} U_{1}^{-1}\nonumber\\
&=&U_{1} \bigg{[}J_{5}^{0}-2 N_{f}(\frac{\alpha_{s}}{4\pi}K^{0}) \bigg{]} U_{1}^{-1}\nonumber\\
&=&U_{1} J_{5}^{0} U_{1}^{-1} - U_{1} 2 N_{f}(\frac{\alpha_{s}}{4\pi}K^{0}) U_{1}^{-1}\nonumber\\
&=&J_{5}^{0}-2 N_{f}(\frac{\alpha_{s}}{4\pi}K^{0})+2 N_{f}\nonumber\\
&=&\widetilde Q_{5}+2 N_{f}~, 
\end {eqnarray}
where in arriving at the last line of the equation use has been made of (\ref{new}). This is a very important 
result as it implies that the $\theta$--vacuum is not invariant under chiral transformations. To see this one computes
\begin {eqnarray}
\label{chiral}
U_{1}\big{[}e^{i \alpha \widetilde Q_{5}}|\theta \rangle \big{]}&=&U_{1} e^{i \alpha \widetilde
Q_{5}} U_{1}^{-1} U_{1} |\theta \rangle\nonumber\\
&=&e^{i(\theta+2N_{f}\alpha)}\big{[}e^{i \alpha \widetilde Q_{5}}|\theta \rangle \big{]}~,
\end {eqnarray}
which says that after a chiral rotation by an angle $\alpha$ the $\theta$ parameter is shifted by $2 N_{f}\alpha$. 
Therefore, similar to the fact that the gauge transformations shift the winding numbers, or equivalently, cause 
transitions between vacua of different homotopy classes, the chiral transformations shift the $\theta$--vacuum depending
on the number of massless quark flavors. This very shift of the $\theta$ parameter under a chiral transformation 
has important implications concerning the quark sector of (\ref{l0lag}).

It has long been known that CP is not a respected symmetry in the neutral $K$--meson ($K^0$) system since the mixings
as well as the decays of these mesons apparently violate the CP symmetry \cite{cr,kaon}. Since the QCD vacuum
angle $\theta$ does not contribute to $K^0$ CP, it is clear that there should be finite CP violation in the electroweak
sector (which will be throughly discussed in the next chapter), or simply, in the quark mass matrices in (\ref{l0lag}).
The sensitivity of the $\theta$--vacuum to the chiral transformations thus implies a hand--shaking of the vacuum angle $\theta$
and the phases of the quark mass matrix ${\cal M}_{q}$.

In general, the mass matrix of quarks is non--Hermitian, so that part of (\ref{l0lag}) containing the quark mass terms 
can be rewritten as  
\begin {equation}
{\cal L}_{M}=\overline{\psi}_{R_{a}}M_{a b}\psi_{L_{b}}+\overline{\psi}_{L_{a}}(M^{\dag})_{a b}\psi_{R_{b}}~,
\end {equation}
where $a, b$ are flavour indices, and $M_{a b}$ are elements of ${\cal{M}}_{q}$. However, by a chiral $SU(N_f)$ 
( here $N_f=3$) transformation 
\begin {equation}
\label{unit}
\psi_{R(L)}^{\prime}=U_{R(L)}\psi_{R(L)},~~~~~~~~~ U_{R(L)} \in SU_{R(L)}(N),
\end {equation}
one can diagonalize ${\cal M}_{q}$ with positive elements up to an overall phase factor. Thus, without loss of 
generality the elements of ${\cal M}_{q}$ can be assumed to have the form 
\begin {equation}
\label{mass}
M_{a b}=m_{a}\delta_{a b}\ e^{i\rho}~,
\end {equation}
where it is clear that the common phase factor here  cannot be rotated away. Using this form of ${\cal{M}}_{q}$ in 
(\ref{l0lag}) one observes that after a further chiral rotation, similar to (\ref{chiral}), by the angle 
$\alpha_0= - \theta/(2 N_{f})$ it is clear that the topological term ${\cal{L}}_{\theta}$ disappears from the 
QCD Lagrangian (\ref{l0lag}). However, this very chiral transformation acts on the quarks as $\psi\rightarrow 
e^{-i \gamma_{5} \alpha_0} \psi$. Therefore, the expression for quark masses (\ref{mass}) now goes over to 
\begin{equation}
\overline{M_{a b}}=m_{a}\delta_{a b}\ e^{i\overline{\rho}}
\end{equation}  
where 
\begin{equation}
\label{finphase}
\overline{\rho}=\rho- \frac{\theta}{N_{f}}
\end{equation} 
is the net phase of the quark mass matrix. One notices that now there is no $\theta$ term in the original QCD Lagrangian (\ref{l0lag}); however,
the same phase is now moved to the quark mass Lagrangian through the angle $\overline{\rho}$. In general the hand--shaking 
between the CP phases in the QCD vacuum ($\theta$) and the phases in the quark mass matrices ($\rho$) is expressed through the effective
$\theta$ angle:
\begin{eqnarray}
\label{tetbar}
\overline{\theta}\equiv - N_{f} \overline{\rho}= \theta - \mbox{Arg}\ \mbox{Det} [{\cal{M}}_{q}]
\end{eqnarray}
where, obviously, $\mbox{Arg}\ \mbox{Det} [{\cal{M}}_{q}]= N_{f}\ \rho$. 
 
Obviously it is the nature of (\ref{tetbar}) which makes the CP violation by strong interactions nontrivial. Indeed, 
if $\theta$ where stable under chiral transformations then there would not be any contribution to $\overline{\theta}$ from
the phases of the quark masses. In such an instance, setting $\theta$ to some value (say, zero) would be a natural 
operation because $\theta$ cannot anyhow be determined from the theory itself. However, even if one chooses $\theta$ 
zero initially, this choice cannot be kept unchanging due to the nonvanishing (because of finite $K^0$ CP) phases of the quark 
mass matrices. Therefore, the CP violation in the QCD Lagrangian (\ref{l0lag}) is not removable at all. One can interpret 
it either as the topological ${\cal{L}}_{\theta}$ piece in (\ref{l0lag}) with real quark masses, or equivalently, as
the quark mass Lagrangian with the phase (\ref{finphase}) with ${\cal{L}}_{\theta}$ absent. Except for this hand--shaking,
there is an unremovable source of CP violation in the QCD Lagrangian with primarily nonperturbative, topological origin.
In the next section we discuss an observable effect of $\overline{\theta}$: the EDM of the neutron.

\section {Computing EDM of Neutron}

As was stated in the Introduction, the nonvanishing, unremovable $\overline{\theta}$ induces a finite EDM for the neutron. 
Calculating this effect with a high degree of precision is difficult as it is a long distance effect, and it may not be 
possible to employ the parton picture. In actuality, one must employ either an empirical model for the neutron structure 
(say, the Bag Model) \cite{bl} or make use of the current algebra techniques \cite{current}. However, the main purpose 
of the work here is to derive the effective Lagrangian ($\delta {\cal{L}}_{CP}$ below) responsible for CP--violating 
effects in the strong interactions. Therefore, instead of a detailed calculation we will quote the results of the earlier
works in estimating the neutron EDM. 

As a starting operation one should first determine the $CP$--violating Lagrangian which is responsible for developing the
EDM. This Lagrangian, $\delta \cal L (CP)$, follows from the quark mass Lagrangian. Consistent with all the above 
calculations we consider three light flavours $u$, $d$ and $s$. As suggested by (\ref{tetbar}) if $\overline \theta$
or at least one of the quark masses vanishes (that is, the determinant itself vanishes) then $\delta {\cal L}(CP) \to 0$.
To make it clear, consider an appropriate unitary transformation reducing the quark mass Lagrangian with the mass term in 
(\ref{mass})  to 
\begin {equation}
{\cal L}_{\tilde M} \longrightarrow {\cal L}_{\tilde M}^{'}=(e^{3 i \overline \rho})m_{u} \overline u_{R} u_{L}+m_{d} \overline
d_{R} d_{L}+m_{s} \overline s_{R} s_{L}+H.C.
\end {equation}
where we have arbitrarily transferred the entire phase content to the $u$ quark mass.  From this we can see that if $m_{u}\to 0$ 
then $\overline \rho$ drops out. Therefore,  $\delta \cal L (CP)$ cannot be simply part of the quark mass Lagrangian having
imaginary parts. As mentioned in the introductory section, the spontaneous break--down of the  axial $SU(3)_A$ symmetry of the QCD 
Lagrangian produces a total of eight Goldstone bosons. In this case the vacuum is infinitely degenerate, and the naive perturbation
theory is not reliable. To determine  $\delta \cal L (CP)$ we have to select one vacuum state out of this infinitely many sets, which 
of course leaves the physical content of the theory unchanged. To leading approximation the vacuum is flavour--blind. The correct
perturbation $\delta\cal L$ should be chosen unitary--equivalent quark mass matrices (\ref{unit}). Moreover, this perturbation
should cause a minimal shift of the vacuum energy with no danger of destabilizing it:
\begin {equation}
\langle \Omega|\delta{\cal L}|\Omega \rangle=\min\limits_{U_{R,L}\in SU(3)_{R,L}}\langle \Omega|{\cal L}_{\tilde
M}(U_{R,L})|\Omega\rangle.
\end {equation}
where $\Omega$ stands for the vacuum. This constrained selection of states with unitarily equivalent quark mass matrices
can be done with the help of Dashen's theorem (\cite{dashen} which assumes that ($i$) the transformations should 
be pure chiral,
$U_{R}^{\dagger}=U_{L}$, and ($ii$) $\delta L$ is flavour--blind. With these conditions at hand, one can parameterize $\delta L$ as 
\begin {equation} 
\label{eqn}
\delta{\cal L}=\overline\psi_{R}^{a}(\mu_{a}+i\omega)\psi_{L}^{a}e^{- i \delta}+H.c.,
\end {equation}
where $\mu_{a}$ and $\omega$ are assumed to be {\it real} parameters and are fixed by\\
\begin{eqnarray}
\label{defin}
&&m_{a}^{2}=\mu_{a}^{2}+\omega^{2},\\
&&\theta^{\prime}=\overline \theta + \delta=\frac{1}{3} \mbox{Arg} \prod \limits_{a}(\mu_{a}+i \omega),\\
&&\mbox{when}\ \  \mu_{a}>0.
\end{eqnarray}

Now we will derive the correct form of $CP$ violating perturbation. Consider (\ref{eqn})
\begin {eqnarray}
\delta{\cal L}&=&\overline \psi^{a}_{R}(\mu_{a}+i\omega)\psi_{L}^{a}e^{- i \delta}+\overline \psi^{a}_{L}
(\mu_{a}-i\omega)\psi_{R}^{a}e^{i \delta}\nonumber\\
&=&\overline\psi^{a}L(\mu_{a}+i\omega)L\psi^{a}e^{- i \delta}+\overline\psi^{a}R(\mu_{a}-i\omega)R\psi^{a}
e^{i \delta}\nonumber\\
&=&\overline\psi^{a}L(\mu_{a}+i\omega)\psi^{a}e^{- i \delta}+\overline\psi^{a}R(\mu_{a}-i\omega)\psi^{a}
e^{i \delta}\nonumber\\
&=&\overline\psi^{a}\big{[}\mu_{a}(L+R)+i\omega(L-R)\big{]}\psi^{a}e^{i \delta}\nonumber\\
&=&\overline\psi^{a}\big{[}\mu_{a}+i\omega\gamma^{5}\big{]}\psi^{a}e^{i \delta}
\end {eqnarray}
where we have used,
\begin{equation}
L=\frac{1+\gamma_{5}}{2},~~~~~~~~~~R=\frac{1-\gamma_{5}}{2}
\end{equation}
Hence $\delta{\cal L}(CP)$ is the $CP$ nonconserving part we need excluding $\delta$. Therefore,
\begin{equation}
\label{lcp}
\delta{\cal L}(CP)=\mp \omega \overline{\psi} i \gamma_{5}\psi
\end{equation}
where $\omega$ can be solved by inverting (\ref{defin}) as:
\begin{equation}
\overline{\theta}=\frac{1}{3}\mbox{\textrm Arg}\bigg{[}(\mu_{1}+i\omega)(\mu_{2}+i\omega)(\mu_{3}+i\omega)\bigg{]}
\end{equation}
therefore 
\begin {eqnarray}
\mbox{\textrm tan}\overline{\theta} &=&\frac{1}{3}\mbox{\textrm tan Arg}\bigg{[}(\mu_{1}+i\omega)
(\mu_{2}+i\omega)(\mu_{3}+i\omega)\bigg{]}\nonumber\\
&=&\frac{1}{3}~~~~\frac{\frac{\omega}{\mu_{1}}+\frac{\omega}{\mu_{2}}+\frac{\omega}{\mu_{3}}+
\frac{\omega^{3}}{\mu_{1}\mu_{2}\mu_{3}}} {1+\omega^{2}(\frac{1}{\mu_{1}\mu_{2}}+
\frac{1}{\mu_{1}\mu_{3}}+\frac{1}{\mu_{2}\mu_{3}})}\nonumber~.\\
\end{eqnarray}
In the limit $|\overline{\theta}|<<1$, and when also $\omega$ is small ( in order to neglect 
$\omega^{2},\omega^{3}$)  we obtain
\begin {equation}
\omega=3~~\frac{\mu_{1}\mu_{2}\mu_{3}}{\mu_{1}\mu_{2}+\mu_{1}\mu_{3}+\mu_{2}\mu_{3}}\overline{\theta}
\end {equation}
From (\ref{defin}) for small $\omega$ we can write $m_{a}^{2}\equiv\mu_{a}^{2}$, which upon substitution in the last 
expression yields,
\begin {equation}
\omega=3~~\frac{m_{1}m_{2}m_{3}}{m_{1}m_{2}+m_{1}m_{3}+m_{2}m_{3}}\overline{\theta}
\end{equation}
which can be used in (\ref{lcp}) to rewrite it as 
\begin {equation}
\delta{\cal L}(CP)=\mp\frac{3 m_{u} m_{d} m_{s}}{m_{u} m_{d}+m_{u} m_{s}+m_{d}
m_{s}}~~~~\overline \theta(\overline \psi i \gamma_{5}\psi)~.
\end{equation}
Derivation of this Lagrangian is the most important step in studying the CP--violating effects in the strong
interactions. It is this very Lagrangian with which there arise CP--violating effects in the hadronic interactions.
The EDM of the neutron is one such fundamental quantity through which the the parameter $\overline\theta$ becomes
observable. Below we will sketch its derivation without going into  details of the hadronic model adopted. Two 
existing calculations due to \cite{bl} and \cite{current} are in good agreement. 

We now choose to sketch the derivation in \cite{current}. The neutron EDM is defined via the correlator
\begin{eqnarray}
\label{corre}
\langle n\left(p+\frac{k}{2}\right)| J_{\mu}^{em}\left(k \right) i \int d^{4} x\ \delta {\cal L}(CP) |
n\left(p-\frac{k}{2}\right)\rangle \equiv -\ D_{n}\left(k^2\right) \overline{u} \sigma_{\mu\nu} k^{\nu} u~,
\end{eqnarray} 
where $J_{\mu}^{em}$ electromagnetic current of the quarks. The value of the formfactor $D(k^2)$ for
on--shell photon (that is, $D(k^2=0)$) is the EDM of the neutron.  The Figure below illustrates the correlator
above. Here the black blobs designate the usual $\pi^{-} n p$ and $\pi^- \pi^+ \gamma$ vertices. The grey blob,
however, shows the CP--violating $\pi n p$ vertex generated by $\delta {\cal{L}}_{CP}$ in the correlator above.
\begin{figure}[ht]
\begin{center}
\includegraphics[angle=0, width=7cm]{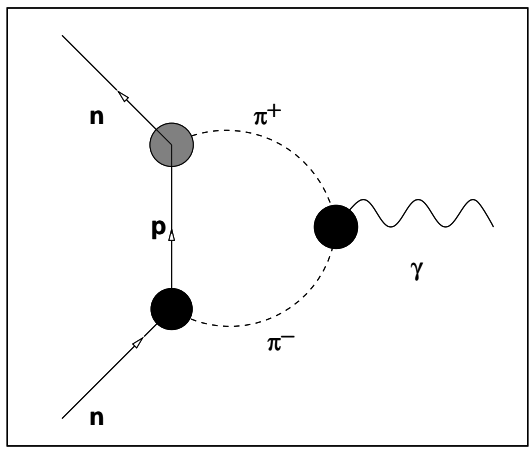}
\caption{The mechanism of generating the neutron EDM. The black blobs are the usual CP--conserving
hadronic couplings. The grey blob shows the CP--violating $\pi^+ n p $ vertex generated by $\delta {\cal{L}}_{CP}$.}
\end{center}
\end{figure}

In general the calculation of the correlator (\ref{corre}) involves a summation over all intermediate states 
$$\sum_{X} \langle n | J_{\mu}^{em} | X\rangle \times  \langle X | \delta {\cal L}(CP) | n\rangle $$ where 
$|X\rangle =  \left\{ |N\rangle,\right.$ $|\pi N\rangle,$ $|\pi \pi N\rangle, \left. \cdots \right\}$. 
However, as shown in \cite{current}, it is the $|\pi N\rangle$ contribution that dominates the amplitude, 
in accordance with the figure above.  The reason for the dominance of this particular intermediate state 
follows from the fact that the neutron disassociates into charged constituents (needed to emit a photon)
in acquiring an EDM. The largest contribution comes from the lightest constituents ( pion) because it 
attains the greatest distance from the center of the charge distribution.
 
Due to the explicit CP violation in the system the pion--nucleon interactions are now generalized to 
\begin{eqnarray}
{\cal{L}}_{\pi N N } = \vec{\pi}\cdot \overline{N} \vec{\sigma} \left( i g_{\pi N N} \gamma_{5} + \overline{g_{\pi N N}}\right) N
\end{eqnarray}
where $\overline{g_{\pi N N}}$ reflects the CP violation.  A direct evaluation of $\langle \pi^{a} N | \delta {\cal{L}}_{CP} | N
\rangle$ leads one to $\overline{g_{\pi N N}}$
\begin{eqnarray}
\overline{g_{\pi N N}}= -\  \overline{\theta}\times  \frac{ M_{\Xi}- M_{N} } {F_\pi}\times \frac{ m_u m_d \left(2 m_s - m_u -
m_d\right)}{m_u +m_d}
\end{eqnarray} 
which is proportional to $\overline{\theta}$ as expected. The remaining vertex in the correlator (\ref{corre}) is 
$\langle n | J_{\mu}^{em} | \pi N\rangle$ is known to be proportional to $\gamma_5 \sigma_{\mu\nu} k^{\nu}$ \cite{gam5}. 

Then a direct computation of the diagram in the figure above gives 
\begin{eqnarray}
d_{n}\equiv D_{n}(k^2=0)= \frac{g_{\pi N N} \overline{g_{\pi N N}}}{4 \pi^{2} M_{N}} \ln \left( \frac{M_{N}}{m_{\pi}}\right)~,     
\end{eqnarray}
which equals numerically $\sim \overline{\theta} \times 2 \times 10^{-16}{\rm e-cm}$ \cite{current}. 

Here one notices that this numerical estimate is quite close to the results reported by Baluni \cite{bl} though two calculations  adopt
different methods in computing the hadronic matrix elements. However, a more interesting coincidence comes by the recent 
estimate \cite{pospelov} of $d_n$ using the QCD sum rules \cite{qcdsumrul}. In this calculation the main object is the
two--point correlator of the neutron current (that current which excites neutron from the QCD vacuum) in that background 
containing the electromagnetic field and the $\theta$--term derived above.

\section{Remarks}
The numerical value of the neutron EDM given above is unfortunately far above the experimental upper bound: $|d_n|\simlt 10^{-25}{\rm
e -cm}$ \cite{exp3}. This means that the effective QCD vacuum angle should satisfy $|\overline{\theta}|\simlt 10^{-9}$ ! This requires 
a huge fine tuning of the pure QCD angle $\theta$ and the phases of the quark mass matrices (\ref{tetbar}). That one has 
such a small value for $\overline{\theta}$ instead of the expected order of unity poses the well--known $strong$ CP $problem$ -- 
a CP hierarchy, or a fine--tuning, or a naturalness problem \cite{strongCP}. 

We will not attempt to discuss possible solutions to this issue instead we will summarize shortly existing proposals. There are
two main ideas towards a solution to the strong CP problem: Relaxation and cancellation. The former refers to the celebrated 
Peccei--Quinn solution \cite{pq} (see also reviews \cite{peccei}) according to which $\overline{\theta}$ (being now a dynamical
variable) relaxes to zero and remains so to all orders in perturbation theory. More explicitly, one 
one ($i$) promotes the phases of the quark mass matrices to dynamical variables (corresponding to the Goldstone 
bosons of spontaneously broken global symmetries), then  ($ii$) computes the instanton--induced effective potential for
$\overline{\theta}$, and finally ($iii$) shows that this potential is actually minimized for $\overline{\theta}=0$. Therefore,
in this picture, $\overline{\theta}$ relaxes to zero dynamically resulting in a vanishing EDM for the neutron. This Peccei--Quinn
idea has generated several other versions in quest for an experimentally viable model: Weinberg \cite{wein} and Wilczek \cite{wil}
put the Peccei--Quinn scale to the weak scale which disagreed with the phenomenological constraints then, later Dine $et. al$ shifted
it to the unification scale \cite{dfsz}. As another version, instead of working with the quark mass matrices directly, Kim and Shifman
$et. al$ introduced additional heavy color triplets \cite{ksvz}. In all versions of the Peccei--Quinn idea the effective vacuum angle
$\overline{\theta}$ possesses a dynamical character and then relaxes to a purely CP--conserving point via the background instanton
effects.   

The other proposal to solve the strong CP problem runs via an appropriate choice of the quark mass matrix ${\cal{M}}_q$
(which  is block diagonal in up-- and down--sector quark mass matrices ) such that $\mbox{Det}[{\cal{M}}_u {\cal{M}}_d]$ is real
whereas the CKM matrix (discussed in the next chapter) keeps having a finite phase \cite{nelson}, as originally proposed by Ann Nelson.
With such a cancellation of the phases coming from the quark mass matrices, the interplay between the original QCD vacuum angle and
that of the quark sector is lost, so that one can take $\theta\equiv 0$ from the scratch. For instance, in the Babu--Mohapatra model \cite{moha}, the theory is
left--right symmetric (parity--conserving) at higher energies so that the topological term is not allowed to contribute at all.
Subsequently a finite $\overline{\theta}$ could be generated were not it for the flavour structure of the quark mass matrices. 

At this point it is convenient to comment on the situation in the supersymmetric models concerning both ideas above. The supersymmetric
models provide enough global symmetries replacing the Peccei--Quinn symmetry together with novel sources of CP violation coming 
from the soft terms \cite{cpphase}. The Peccei--Quinn solution can therefore be adapted to supersymmetry for not only solving the
strong CP problem but also solving its own hierarchy problems \cite{gluino,summ}. Finally, the Peccei--Quinn idea can be generalized 
also for examining the CP hierarchy problem of supersymmetry in a dynamical way \cite{dyna}. Apart from the operation of the
Peccei--Quinn mechanism in supersymmetry, it is possible to implement supersymmetric falvour models which incorporate the Nelson--Barr
models \cite{barr} which can be particularly useful for having observable supersymmetric CP violation \cite{observe}.

\chapter{The Neutron EDM: Electroweak Interactions}
In the framework of the Standard Model (SM) of electroweak and strong interactions the other source of CP 
violation is the single complex phase in the Cabibbo--Kobayashi--Maskawa (CKM) matrix \cite{km,cabibbo}. 
To examine it's consequences for EDM of the neutron we will first briefly review the electroweak sector of the SM 
with particular emphasis on the CP--violation currents and parameterizations of the CKM matrix.

\section {Electroweak Lagrangian and the CKM matrix}
The particle spectrum of the SM consists of gauge and Higgs bosons together with three families of quarks and leptons (See \cite{ali}
for a review):
\\
\underline{Fermions}
\begin{displaymath}
\mathbf{Leptons}:
\left( \begin{array}{c}
\nu_{e}\\
e\\
\end{array} \right)_{L},~~~~~~~\left( \begin{array}{c}
\nu_{\mu}\\
\mu
\end{array} \right)_{L},~~~~~~\left( \begin{array}{c}
\nu_{\tau}\\
\tau
\end{array} \right)_{L};~~~e_{R}, \mu_{R}, \tau_{R}
\end{displaymath}
\begin{displaymath}
\mathbf{Quarks}:
\left( \begin{array}{c}
{\it u}\\
{\it d}
\end{array} \right)_{L},~~~~~~~\left( \begin{array}{c}
{\it c}\\
{\it s}
\end{array} \right)_{L},~~~~~~\left( \begin{array}{c}
{\it t}\\
{\it b}
\end{array} \right)_{L};~~~u_{R}, d_{R}, c_{R},\ldots
\end{displaymath}
\underline{Gauge Bosons}
\begin{displaymath}
\left( \begin{array}{ccc}
W_{\mu}^{1} & W_{\mu}^{2} & W_{\mu}^{3}
\end{array} \right);~~~~~~~B_{\mu}
\end{displaymath}
\underline{Scalars}
\begin{equation}
\Phi=
\left( \begin{array}{c}
\phi^{+}\\
\phi^{0}\\
\end{array} \right),~~\Phi^{\dagger}=\left( \begin{array}{c}
\phi^{-}\\
{\phi^{\ast}}^{0}\\
\end{array} \right)
\end{equation}
The interaction between the fermions and the gauge bosons has the form:
\begin{equation}
\label{ewlag}
{\cal L}(f, W, B)=\sum\limits_{j=1}^{3}\big{\{} \bar{l}^{j}_{L} D\!\!\!\!\!\diagup l^{j}_{L}+\bar{l}^{j}_{R} D'\!\!\!\!\!\!\!\diagup
l^{j}_{R}+\bar{q}^{j}_{L} D\!\!\!\!\!\diagup q^{j}_{L} \big{\}}+\sum\limits_{i=1}^{6}\bar{q}^{i}_{R} D'\!\!\!\!\!\!\!\diagup q^{i}_{R}
\end{equation}
where j is the family index,
\begin{displaymath}
l_{L}^{1}=
\left( \begin{array}{c}
\nu_{e}\\
e
\end{array} \right)_{L},~~~l_{R}^{1}=e_{R},~~~~~~~\left( \begin{array}{c}
{\it u}\\
{\it d}
\end{array} \right)_{L},~~~q_{R}^{1}=u_{R},~~q_{R}^{2}=d_{R},\ldots
\end{displaymath}
and the two covariant derivatives are defined as:
\begin{eqnarray}
D\!\!\!\!\!\diagup \equiv D_{\mu}\gamma^{\mu},~~~D'\!\!\!\!\!\!\!\diagup \equiv D'_{\mu}\gamma^{\mu} \nonumber\\
D_{\mu}=\partial_{\mu}-i g \bigg{(} \vec{W}_{\mu}.\frac{\vec \sigma}{2} \bigg{)}-i g_{1}\frac{Y}{2} B_{\mu} \nonumber\\
D'_{\mu}=\partial_{\mu}-i g_{1}\frac{Y}{2} B_{\mu}.
\end{eqnarray}
where $g_{1}$ and $g_{2}$ are respectively, the $U(1)_{Y}$ and $SU(2)$, coupling constants, and $\sigma^{a} (a=1, 2, 3)$ are the
(weak) isospin matrices.
The interaction term ${\cal L}(f,\phi)$ involving the fermions and the Higgs fields has the Yukawa form,
\begin{equation}
\label{lyuk}
{\cal L}(f, \phi)=\sum\limits_{j=1}^{3} \big{\{} (h_{i})_{j} \bar l^{j}_{L} \Phi l^{j}_{R} \big{\}}+\sum\limits_{j, k=1}^{3} \bigg{\{}
( h'_{q})_{jk} \bar q^{j}_{L} \Phi u_{R}^{k}+(h_{q})_{jk} \bar q_{L}^{j} \Phi^{c} d_{R}^{k} \bigg{\}}
\end{equation}
where the charge--conjugated scalar doublet reads as
\begin{displaymath}
\Phi^{c}=i \sigma_{2} \phi^{*}=
\left( \begin{array}{c}
\phi^{0*}\\
- \phi^{-}
\end{array} \right)
\end{displaymath}
with both $\Phi$ and $\Phi^{c}$ transforming as a (weak) isospin doublet with opposite hypercharges.\\
After the spontaneous symmetry breaking $SU(2) \otimes U(1)\longrightarrow U(1)_{EM}$, the gauge bosons, fermions and the neutral
scalar field, $\phi$ , acquire non-zero masses through the Higgs mechanism. After $\phi$ develops a vacuum expectation value $v$,
one can expand the Higgs doublet around this ground state so that (\ref{lyuk}) becomes:
\begin{equation}
\label{ssb}
{\cal L}(f, \phi)^{SSB}=\sum\limits_{j=1}^{3}(m_{j})_{l}\bar l^{j}_{L}l^{j}_{R}\bigg{(}1+\frac{1}{\nu}\phi \bigg{)}-\sum\limits_{j,
k=1}^{3}\big{\{} (m_{jk})_{U}\bar u_{L}^{j}u_{R}^{k}+(m_{jk})_{D}\bar d_{L}^{j}d_{R}^{k} \bigg{\}}\bigg{(}1+\frac{1}{\nu}\phi
\bigg{)}+h. c.
\end{equation}
where the fermions mass matrices (in the flavour space) are defined by 
\begin{eqnarray}
(m_{j})_{l}=(h_{j})_{l}\frac{\nu}{\sqrt 2}\nonumber\\
(m_{jk})_{U}=-(h_{q})_{jk}\frac{\nu}{\sqrt 2}\nonumber\\
(m_{jk})_{D}=-(h'_{q})_{jk}\frac{\nu}{\sqrt 2}~,
\end{eqnarray}
with  $(m_{jk})_{U}$ and $(m_{jk})_{D}$ being $3\times  3$ mass matrices for up and down quarks, respectively. In order to write
the Lagrangian in terms of quark mass eigenstates, the mass matrices $(m_{jk})_{U}$ and $(m_{jk})_{D}$ have to be diagonalized. This
can be done with the help of two unitary matrices usually denoted by $V_{L}^{up}$ and $V_{R}^{up \dagger}$ (similarly for down 
quarks):
\begin{eqnarray}
V_{L}^{up} m_{U} V_{R}^{up \dagger}\equiv (m_{diag.})_{U}\equiv Diag.(m_{u}, m_{c}, m_{t})\nonumber\\
V_{L}^{down} m_{D} V_{R}^{down \dagger}\equiv (m_{diag.})_{D}\equiv Diag.(m_{d}, m_{s}, m_{b})
\end{eqnarray}
with $V_{L}^{up \dagger} v_{L}^{up}=1$ , etc. Considering only up quarks we can write the mass term as:
\begin{eqnarray}
\label{2.7}
\bar u_{L} m_{U} u_{R}&=&\bar u_{L}V^{up \dagger}_{L} V_{L}^{up} m_{U} V_{R}^{up \dagger} V_{R}^{up} u_{R}\nonumber\\
&=&\overline{u_{L}V_{L}^{up}}(m_{diag.})_{U}(V_{R}^{up} u_{R}).
\end{eqnarray}
which shows that the physical quark states are:
\begin{eqnarray}
\label{2.8}
u_{L}^{Phys}=V_{L}^{up}u_{L}=V_{L}^{up} \left( \begin{array}{c}
u_{L}\\
c_{L}\\
t_{L}\\
\end{array} \right)\nonumber\\
d_{L}^{Phys}=V_{L}^{down}d_{L}=V_{L}^{down} \left( \begin{array}{c}
d_{L}\\
s_{L}\\
b_{L}\\
\end{array} \right)
\end{eqnarray}
Then in terms of the mass eigenstates, (\ref{ssb}) can be rewritten as:
\begin{equation}
\label{2.10}
{\cal L}(f, \phi)^{SSB}=-\bigg{(}1+\frac{1}{\nu}\phi \bigg{)} \bigg{\{}\sum\limits_{i=1}^{6}m_{q_{i}}\bar q_{i}
q_{i}+\sum\limits_{j=1}^{3}m_{l_{j}}\bar l_{j} l_{j} \bigg{\}}
\end{equation}
The identification of the parameters $m_{l_{i}}, m_{q_{i}}$ with the lepton and quark masses is now clear. Since we have written
the term ${\cal L}(f, \phi)$ in terms of the physical quark fields, we will express other terms of the Electroweak Lagrangian in
terms of the physical quark fields like ${\cal L}(f, W, B)$.\\
\\
Considering first the `neutral current', we see that neutral current part of ${\cal L}(f, W, B)$ is manifestly  flavor diagonal.
Written in terms of physical boson $(W_{\mu}^{\pm}, Z_{\mu}^{0}, A_{\mu})$ and fermion fields:
\begin{equation}
\label{2.11}
J_{\mu}^{NC}=\sum\limits_{i}\bar f_{i}\bigg{[} \frac{e}{2 \sqrt{2}\sin^{2}\theta_{W}} Z_{\mu} (I_{3 L}-Q
\sin^{2}\theta_{W})_{i}+e A_{\mu}Q_{i}\bigg{]}f_{i}
\end{equation}
The neutral current interaction induced by the Z-exchange violates P and C but conserves CP.\\
\\
It is important to emphasize here that the Higgs-fermion Yukawa couplings are flavour diagonal so there are no
flavour changing neutral currents (FCNC). Thus, all the flavour changing transitions in the Standard Model are confined to the
charged current (CC) sector.  From (\ref{ewlag}), concentrating only on the quark sector, we obtain the charged current
interaction lagrangian,
\begin{equation}
{\cal L}^{CC}=\frac{e}{2 \sqrt{2} \sin^{2}\theta_{W}}\sum\limits_{i=1}^{3}\bar u^{i}_{L}\gamma^{\mu}W_{\mu}^{+}d_{L}^{i}+h.c.
\end{equation}
Making use of (\ref{2.8}) in (\ref{2.11}) we finally get:
\begin{eqnarray}
{\cal L}^{CC}&=&\frac{e}{2 \sqrt{2}\sin^{2}\theta_{W}}\sum\limits_{i=1}^{3}\bar u^{i}_{L}V_{L}^{up
\dagger}V_{L}^{up}\gamma^{\mu}W_{\mu}^{+}V_{L}^{down \dagger}V_{L}^{down}d_{L}^{i}+h.c \nonumber\\
&=&\frac{e}{2 \sqrt{2}\sin^{2}\theta_{W}}\sum\limits_{i, j=1}^{3}(\bar u^{Phys}_{L})^{i}\gamma^{\mu}W_{\mu}^{+}(V_{L}^{up}V_{L}^{down
\dagger})_{i j}(d_{L}^{Phys})^{j}+h.c. 
\end{eqnarray}
Therefore, the charged current $J_{\mu}^{CC}$ which couples to the $W^{\pm}$ is,
\begin{equation}
J_{\mu}^{CC}=
\left( \begin{array}{ccc}
\bar u, & \bar c, & \bar t\\
\end{array} \right)\gamma_{\mu}V_{CKM}
\left( \begin{array}{c}
d\\
s\\
b\\
\end{array} \right)_{L}
\end{equation}
where $V_{CKM}=V_{L}^{up}V_{L}^{down \dagger}$ is the CKM matrix \cite{km,cabibbo}. It is a generalization of the Cabibbo rotation
\cite{cabibbo} to three quark flavours, and it was introduced to keep the unitarity of weak interactions. The
charged current Lagrangian has a (V-A) structure, hence it violates P and C maximally.\\ 
\\
In general, ${\cal L}^{CC}$ violates CP due to the possibility of some non-trivial phase in $V_{CKM}$. For instance,
consider (1.12), dropping the superscript {\it Phys} and using the CP properties of fermionic fields we get:
\begin{eqnarray}
\label{2.15}
{\cal L}^{CC}&=&\frac{e}{2 \sqrt{2} \sin^{2}\theta_{W}}\sum\limits_{i=1}^{3}\bar u^{i}_{L}\gamma^{\mu}W_{\mu}^{+}(V_{CKM})_{i
j}d_{L}^{j}+\bar d_{L}^{j} \gamma^{\mu} W_{\mu}^{-} (V^{*}_{CKM})_{i j}u^{i}_{L}\nonumber\\
&\overset{CP}\longrightarrow& \bar{d}_{L}^{j}\gamma^{\mu}W_{\mu}^{-}(V_{CKM})
_{ij}u^{i}_{L}+\bar u^{i}_{L}\gamma^{\mu}W_{\mu}^{+}(V^{*}_{CKM})_{ij}
d_{L}^{j} 
\end{eqnarray}
where we have used $W_{\mu}^{+}\overset {CP}\longrightarrow {-W_{\mu}^{-}}$.\\
\\
Thus the CP conservation requires the matrix V to be real. It is real for two families, and hence, for two families of quarks CP is
automatically conserved. But in case of three families of quarks, as was suggested by Kobayashi and Maskawa, the CP is violated
due to the presence of complex phase in $V_{CKM}$.\\
\\
$V_{CKM}$ being a unitary matrix satisfies, $V_{CKM}^{\dagger}V_{CKM}=1$. The matrix elements of $V_{CKM}$ are determined by the
charged current coupling to the $W^{\pm}$ bosons. Symbolically this matrix can be written as:
\begin{equation}
V_{CKM}=
\left( \begin{array}{ccc}
V_{ud} & V_{us} & V_{ub} \\
V_{cd} & V_{cs} & V_{cb} \\
V_{td} & V_{ts} & V_{tb}
\end{array} \right)
\end{equation}
As long as the unitarity is maintained, one can parameterize the CKM matrix using three Euler angles (appropriate for $3\times3$
matrices) and a complex phase $\delta_{CKM}$. There are different parameterizations of CKM matrix. The original one, due to Kobayashi
and Maskawa \cite{km} was constructed from the rotation matrices in the flavour space involving the angles $\theta_{i}(i=1, 2, 3)$ and
the phase $\delta_{CKM}$,
\begin{equation}
V_{KM}=R_{23}(\theta_{3}, \delta_{CKM})R_{12}(\theta_{1}, 0)R_{23}(\theta_{2}, 0)
\end{equation}
where $0 \leq \theta_{i}\leq \frac{\pi}{2},\ \  0\leq \delta_{CKM} \leq 2 \pi$, and $R_{ij}(\theta, \phi)$ denotes a unitary rotation
in the
$(i, j)$ plane by the angle $\theta$ and the phase $\phi$. Then a possible representation for $V_{KM}$ is:
\begin{equation}
V_{KM}=
\left( \begin{array}{ccc}
c_{1} & -s_{1}c_{3} & -s_{1}s_{3}\\
s_{1}c_{2} & c_{1}c_{2}c_{3}-s_{2}s_{3}e^{i \delta_{CKM}} & c_{1}c_{2}s_{3}+s_{2}c_{3}e^{i \delta_{CKM}}\\
s_{1}s_{2} & c_{1}s_{2}c_{3}+c_{2}s_{3}e^{i \delta_{CKM}} &
c_{1}s_{2}s_{3}-c_{2}c_{3}e^{i \delta_{CKM}}
\end{array} \right)
\end{equation} 
with $c_{i}=\cos \theta_{i}, s_{i}=\sin \theta_{i}$. This reduces to the usual Cabibo form for $\theta_{2}=\theta_{3}=0$ with
$\theta_{1}\theta_{c}$ identified with the Cabibo angle.\\
\\
However, in discussing several flavour--changing processes it is useful to consider an approximate form of $V_{KM}$. Indeed, 
Wolfenstein \cite{wolf}, has made made the important observation that (empirically) the $V_{KM}$ can be expressed as
\begin{eqnarray}
|V_{ii}|\simeq~1,~~ i=1\ldots 3 \nonumber\\
|V_{12}|\simeq~|V_{21}|\sim\lambda \nonumber\\
|V_{23}|\simeq~|V_{32}|\sim\lambda^{2} \nonumber\\
|V_{13}|\simeq~|V_{31}|\sim\lambda^{3}
\end{eqnarray}
with $\lambda\equiv \sin \theta_{c}=0.221$. With this hierarchy of elements in terms of the Cabibo angle one can write
\begin{equation}
V_{Wolfenstein}=
\left( \begin{array}{ccc}
1-\frac{1}{2}\lambda^{2} & \lambda & A\lambda^{3}(\rho-i \eta)\\
- \lambda & 1-\frac{1}{2}\lambda^{2} & A\lambda^{2}\\
A\lambda^{3}(1-\rho-i \eta) & -A\lambda^{2} & 1\\
\end{array} \right)
\end{equation}
where $A\sim \rho \sim 1$ and $|\eta|< < 1$. This form of the CKM matrix is rather common so that the improvements 
in the experimental determinations are generally expressed in terms of the parameters here.
Having briefly reviewed the particle content of the Electroweak Theory and the CKM matrix we are now in a position to describe the 
consequences of CP violation in Electro-weak sector for the EDM of neutron.\\

\section{Computation of Neutron EDM and Form Factors}
In the sections which follow we will describe the calculation of the neutron EDM in the electroweak sector. It is clear that 
the end result will be proportional to $\delta_{CKM}$, the only source of CP violation in the electroweak theory. In estimating the
neutron EDM we adopt the quark model where the neutron is composed of three valence quarks ${\bf u}$, ${\bf d}$m and ${\bf d}$. In
principle, if one calculates the EDM of a free quark $q$ at the W--boson mass level then it is a straightforward issue to move it down to
the nucleon mass level using the appropriate renormalization group equations (RGE). Then a rough expression for $d_n$ in terms 
of the quark EDM's reads as 
\begin{eqnarray}
d_n\sim \frac{4}{3} \left(d_{d}\right)_{M_N}-\frac{1}{3}\left(d_u\right)_{M_N}
\end{eqnarray}
where the subscript $M_N$ shows that these moments are calculated at the nucleon mass scale. In estimating the 
contributions of the physics beyond the SM ($e.g.$ supersymmetry \cite{susyEDM} or two--scalar doublet models \cite{2hdmEDM})
this relation works well, and puts stringent bounds on the CP violation sources in the underlying model. Here one
notices that the unknown parameters of the two--doublet models and the CP hierarchy problem in the supersymmetry
can be avoided (at least partially) in models where supersymmetry is broken around $a$ ${\rm TeV}$ leaving a
two--scalar doublet model with less unknowns and appropriate CP violation sources \cite{2hdm2,cpphase}.

\begin{figure}[ht]
\label{21}
\begin{center}
\includegraphics[angle=0, width=14cm]{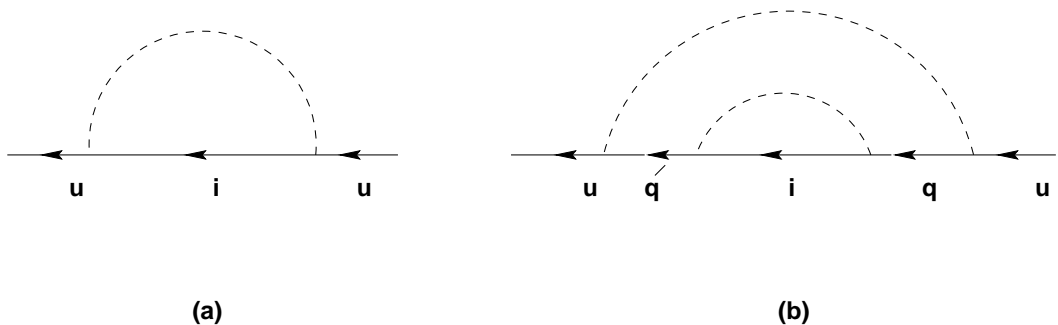}
\caption{Diagrams for induced u-quark dipole moment:the photon line is to be attached to all the charged lines in (a) one-loop and
(b) two-loop diagrams.}
\end{center}
\end{figure}

However, in estimating the contributions of $\delta_{CKM}$ to the neutron EDM one faces with certain difficulties associated
with the EDM of a single free quark. After a detailed repetition of Ellis--Gaillard--Nanopoulos \cite{egn} analysis Shabalin
\cite{shabalin} has shown that there is no contribution to a quark EDM up to three loops. This arises mainly from the cancellations among 
different diagrams due to the unitarity of the CKM matrix. To illustrate the situation it is convenient to refer to Fig. 2. 1. The
one--loop graph (graph (a) here) in this figure is a self--conjugated graph which cannot contribute to CP violation. At the two loop
level, there is a possibility of obtaining ``non-self'' conjugate diagrams, as seen from Fig. 2. 1. (b). Indeed, this diagram yields a
vertex of the form 
\begin{equation}
\label{edm}
e d_{n}(k^{2})\overline{u}\sigma_{\mu \nu}k^\nu \gamma_5 u A^\mu
\end{equation}
where $d_{n}(k^{2})$ in the limit $k^2\longrightarrow 0$ is the EDM, and is suppressed by the usual GIM factor of
$\Delta m^{2}/M_{W}^{2}=\sum\limits_{i} V_{q i}V_{q^{'} i}^{*}m_{i}^{2}M_{W}^{2}$. Unlike the expectations of \cite{egn}, it was shown 
by \cite{shabalin} that these two--loop contributions sum up to zero.  Hence one is left with the
three--loop and higher contributions for the EDM of a single free quark though even the three--loop
contributions were already shown to partially cancel \cite{acbk}. These cancellations bring the d-quark contribution
down to $10^{-34}$ e-cm.

After illustrating the smallness of the single quark EDM's with electroweak theory, we note that the problem here can be circumvented
through a more realistic approach to the problem. Indeed, one has to take into account the multi--parton content of the neutron in
calculating the EDM, that is, instead of dealing with a single free quark, one should consider all three quarks simultaneously. The
original analysis by  Nanopoulos--Y{\i}ld{\i}z--Cox \cite{nn*} has thus dealt mainly with the diquark moments, where the third valence
quark was taken as the "spectator" of the process in charge of balancing the spin, charge and the kinematics of the nucleon. Therefore, 
below we shall be analyzing the neutron EDM (in parton language) taking into account two (very weakly interacting) quarks into account. 
It will be seen that with such a multi--parton language the nucleon's moment turns out to be essentially a one--loop effect.

\subsection{Diquark Interaction and Neutron Moment}
\begin{figure}[ht]
\label{22}
\begin{center}
\includegraphics[angle=270, width=9cm]{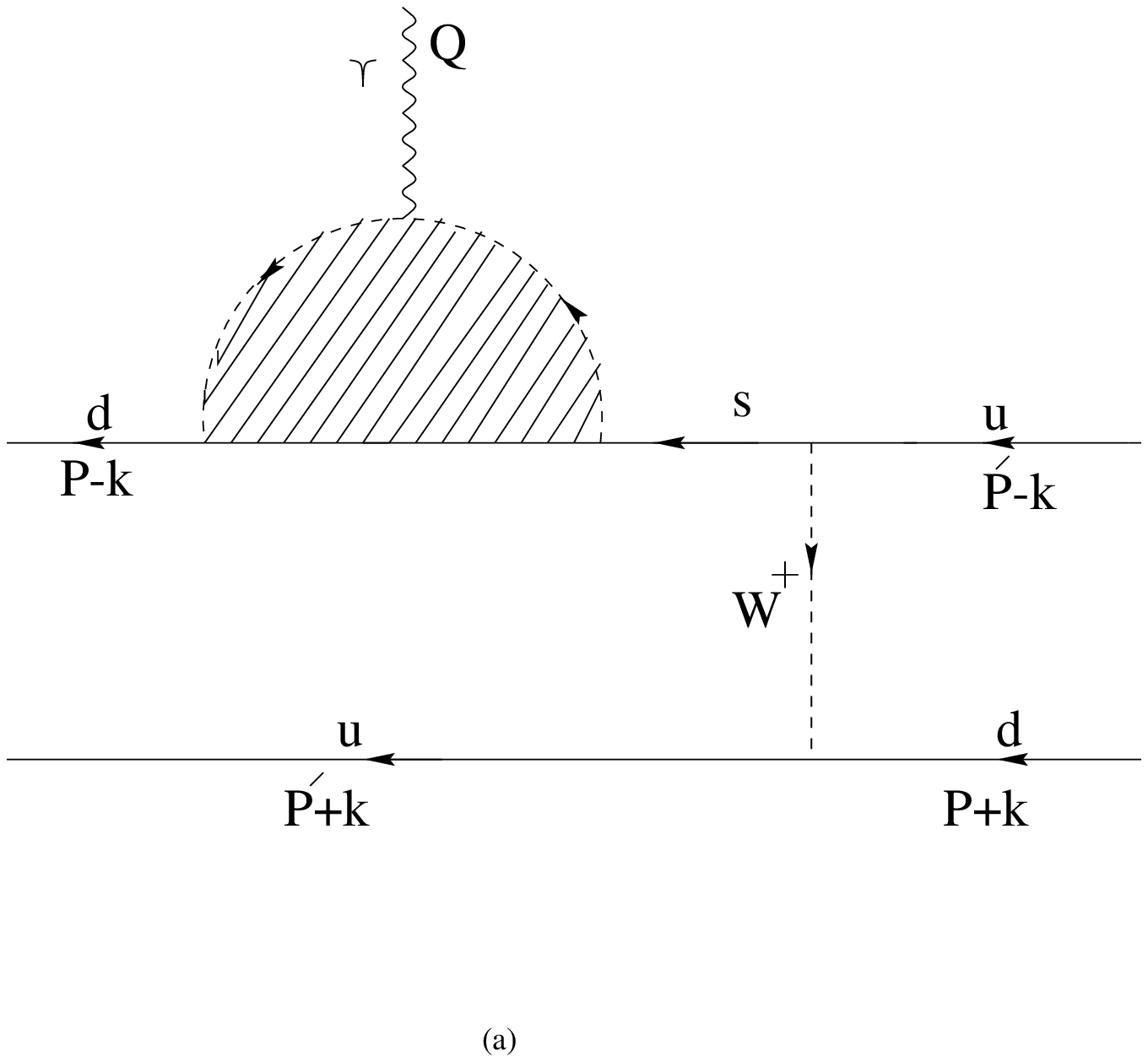}
\caption{Type of diagram contributing to the EDM. Quark lines are labelled according to leading order
contribution; external lines are labelled with momenta ($Q<<P, P^{\prime}, k$).}
\end{center}
\end{figure}
Depicted in Fig. 2. 2 are the class of diagrams contributing to the EDM of a two--quark system where the interaction 
between the two quarks is "weak", that is, mediated by the $W^{\pm}$ boson. The shaded blob in this
figure is the $quark$--$quark$--$photon$ vertex which can get contributions from three possibilities shown in 
Fig. 2. 3. One keeps in mind that $W^{\pm}$ and $\gamma$ can be attached to any of the four external legs of the quarks. 
\begin{figure}[ht]
\label{3point}
\begin{center}  
\includegraphics[angle=0, width=14cm]{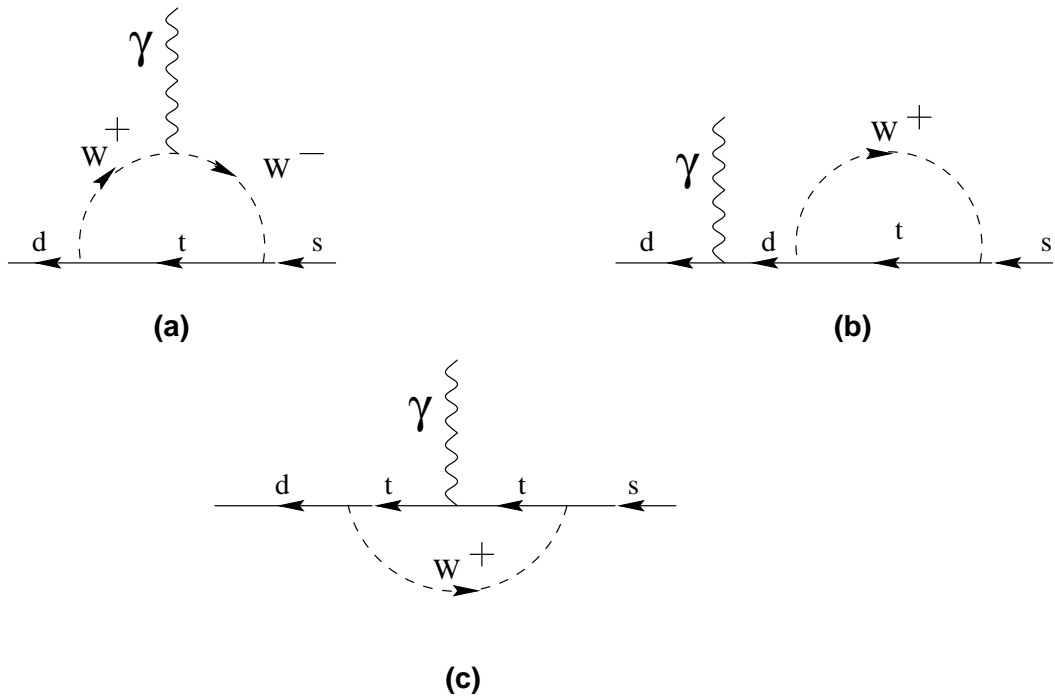}
\caption{The three contributions to the three-particle vertex of \ref{22}}
\end{center}
\end{figure}
Schematically, the amplitudes in Fig. 2. 3 add to give a transition amplitude of the form 
$$\left(\mbox{a form factor}\right)\times \left( Q\!\!\!\!\!\diagup \gamma^{\mu} p\!\!\!\!\!\diagup
- p\!\!\!\!\!\diagup \gamma^{\mu}  Q\!\!\!\!\!\diagup\right) P_{L}$$
where $p_{\mu}$ is the quark momentum, and $Q_{\mu}$ is the momentum of photon of which the quadratic 
terms have been neglected. Direct compoutation yelds a five--particle amplitude
\begin{eqnarray} 
\label{tds}
\Gamma^{\mu}_{dts}&=&\left((8 G_{F}/\sqrt{2}) M_{W}^{2}\right)^{2}~V~F(m_{d}^{2}, m_{t}^{2})\ \bar{d}(P-k)\nonumber\\
&\times&\left[Q\!\!\!\!\!\diagup \gamma^{\mu} (P\!\!\!\!\!\diagup - k\!\!\!\!\!\diagup)-(P\!\!\!\!\!\diagup -
k\!\!\!\!\!\diagup)\gamma^{\mu} Q\!\!\!\!\!\diagup \right]\nonumber\\
&\times&P_L \frac{(P\!\!\!\!\!\diagup -
k\!\!\!\!\!\diagup)+m_{s}}{m_{d}^{2}-m_{s}^{2}}\gamma^{k} P_{L} u(P^{\prime}-k)\nonumber\\
&\times&\Delta_{k \lambda}(P-P^{\prime})\overline{u}(P^{\prime}+k)\gamma^{\lambda} P_{L} d(P+k),
\end{eqnarray}
where $\Delta_{k \lambda}$ is the $W^{\pm}$ propagator in the unitary gauge, and 
$V=(V^{\dagger})_{d t}V_{ts}(V^{\dagger})_{s u}V_{ud}$ is a reparametrization invariant combination of
the relevant elements of the CKM matrix. The quantity $F$ is the form factor, which will be detailed below.

The EDM is given by the coefficient of $\Gamma^{0}_{dts}\sim \vec{ Q}A^0-Q^0 \vec{A}$,
so only $\Gamma^{0}_{dts}$ needs be determined for calculating the diquark EDM. To continue the analysis 
we need to determine first the quark spinors in the diquark rest frame. Parameterizing $\gamma^{\mu}$ as 
\begin{eqnarray}
\gamma^{\mu}&=&\bigg{[}\left ( \begin{array}{cc}
1 & 0\\
0 & -1
\end{array}\right ),
\left ( \begin{array}{cc}
0 & \vec{\sigma}\\
-\vec{\sigma} & 0
\end{array}\right ) \bigg{]}
\end{eqnarray}
the quark spinor ${\bf q}=( {\bf u},~ \mbox{or},~ {\bf d})$  can be written (as  two--component spin states in  non-relativistic 
approximation)
\begin{eqnarray}
{\bf q}(p)&=&[2 m (p^{0}+m)]^{-1/2}
\left ( \begin{array}{c}
p^{0}+m\\
\vec{p}.\vec{\sigma}
\end{array}\right )
q_{\sigma}
\end{eqnarray}
where $q_{\sigma}=\left(\begin{array}{c c} 1\\ 0\end{array}\right)~\ \ \mbox{or}~\ \ \ \left(\begin{array}{c c} 0\\ 
1\end{array}\right)$ are the usual two--component basis vectors. 
Using this construction in the rest frame of the diquark together with an averaging over the orientation of 
relative momentum of the quarks, we obtain the mean value of the five--particle vertex (\ref{tds}) as
\begin{equation}
\label{gamma}
\langle \Gamma^{0}_{dts}\rangle=\frac{1}{3}\bigg{(}\frac{8 G_{F}}{\sqrt{2}}M_{W}^{2}\bigg{)}^{2}~V~\frac{F(m_{d}^{2},
m_{t}^{2})}{m_{d}^{2}-m_{s}^{2}}~~\frac{1}{M_{W}^{2}}K \vec{ Q}.\vec{\sigma}_{n}~.
\end{equation}
In deriving this equation we have made use of the spherical symmetry of the nucleon wavefunction so as to 
have the neutron spin function  
\begin{eqnarray}
\label{neut}
|n,\pm\rangle=\pm\frac{1}{\sqrt{6}}\left\{ |\pm\rangle_{\bf u} \left(|+ - \rangle + |- +\rangle\right)_{\bf d d}
- 2 |\mp\rangle_{\bf u} |\pm \pm\rangle_{\bf d d}\right\}
\end{eqnarray}
after using the appropriate Clebsch--Gordon coefficients. Between such states the evaluation of the spin--dependent
operators can be done using 
\begin{eqnarray}
\vec{Q}\cdot\vec{s_1}-\vec{Q}\cdot\vec{s_2}=\left\{\begin{array}{c c} -\vec{Q}\cdot\vec{\sigma}_{n}~, \mbox{for {\bf u}}\\
+\vec{Q}\cdot\vec{\sigma}_{n} ~, \mbox{for {\bf d}}\end{array}\right.
\end{eqnarray}
where $\vec{\sigma}_n$ are the Pauli matrices for the neutron spin, defined in the basis (\ref{neut}). Another 
quantity in (\ref{gamma}) to be mentioned is $K=E \vec{p}^{2}/m$ involves energy, momentum and mass of either quark
in the diquark rest frame (remember we are using $m_u=m_d\equiv m$).
 
In addition to Fig. 2. 2. there is another diagram with the same structure, however, occuring at the other ${\bf d}$ quark leg.
This diagram gives the same result except for the fact that $V$ is now replaced by the $V^{\ast}$, and the overall 
sign is changed. So the net result will be given by (\ref{gamma}) with $V$ replaced by $2\ i \ \textrm{Im} V$, i.e.
\begin{equation}
\langle \Gamma^{0}_{dts}\rangle=\frac{1}{3}\bigg{(}\frac{8 G_{F}}{\sqrt{2}}M_{W}^{2}\bigg{)}^{2}~2 i \textrm{Im} V~
\frac{F(m_{d}^{2}, m_{t}^{2})}{m_{d}^{2}-m_{s}^{2}}~~\frac{1}{M_{W}^{2}}K \vec{Q}.\vec{\sigma}_{n}~.
\end{equation}
Integrating this result over the quark phase space, supplying a factor of 2 for the ${\vec d}$ quarks, we determine the
appropriate coefficient to be identified with the EDM of the neutron 
\begin{equation}
\label{maim}
d_{n}=\frac{2}{3}~\bigg{(}\frac{8 G_{F}}{\sqrt{2}}\bigg{)}~~(2 i \textrm{Im} V)~\frac{M_{W}^{2}F(m_{d}^{2}, m_{t}^{2})}
{m_{d}^{2}-m_{s}^{2}}~\frac{m_{n}}{\pi}~\Lambda,
\end{equation}
where $\Lambda=\langle m_q |\vec{P}_q|^{3}\rangle$ in the neutron state to be evaluated in the diquark rest frame. 
One notes that $\Lambda\rightarrow 0$ for both $m_q\rightarrow 0$ (ultrareletivistic limit) and $m_{n}-3 m_q\rightarrow 0$
(nonrelativistic limit). Assuming that averaging in the neutron state is given by the phase space alone, $\Lambda$ is
a bounded function of $m_q$ whose maximum being in between the two asymptotics lets one to make the rough estimate 
$\Lambda\simlt
10^{-3}~m_{n}^{4}$. Now we discuss the behaviour of the form factor $F$. It was introduced in (\ref{tds}), and reads as 
\begin{eqnarray}
\label{ff}
F(m_{d}^{2}, m_{t}^{2})&=&\frac{- i}{(4 \pi)^2}~\frac{1}{2 M_{W}^{2}}~\bigg{\{}e \bigg{(}\frac{3}{2}~\frac{m_{t}^2}
{M_{W}^{2}}\bigg{)}-\frac{3}{2}e \bigg{(}-\frac{m_{t}^2}{M_{W}^{2}}\bigg{)}+\frac{2}{3}e \bigg{(}2\frac{m_{t}^2}
{M_{W}^{2}}\bigg{)}\bigg{\}}\nonumber\\
&=&\frac{- i}{(4 \pi)^2}\frac{19}{12}e~\frac{1}{M_{W}^{2}}~\frac{m_{t}^2}{M_{W}^{2}},
\end{eqnarray}
where the three terms in the bracket correspond to the diagrams of Fig. 2. 3. From the experimental values 
of the top quark and $W^{\pm}$ mass we see that the terms in the parenthesis equals numerically $\sim 10$.
It is clear that, if the top quark were not heavy the ratio $m_t^2/M_{W}^{2}$ would have brought about a GIM 
suppression factor, as usual.  

Using the parameterization of the CKM matrix introduced in the previous section we can write 
\begin{equation}
\textrm{Im}\ V=c_{1}s_{1}^{2}c_{2}s_{2}c_{3}s_{3} \sin\delta_{CKM}, 
\end{equation}
with $c_{i}=\cos \theta_{i}, s_{i}=\sin \theta_{i}$, and the angles $\theta_{i}$ and $\delta_{CKM}$ are the parameters of 
CKM matrix.

Substituting (\ref{ff}) into (\ref{maim}), the neutron EDM turs out to have the following explicit expression
\begin{eqnarray}
\label{edmn}
d_{n}&=&\frac{2}{3}~\bigg{(}\frac{8 G_{F}}{\sqrt{2}}\bigg{)}~~(2 i \textrm{Im} V)~\frac{- i}{(4 \pi)^2}\frac{19}{12}e~
\frac{1}{M_{W}^{2}}~\frac{m_{t}^2}{M_{W}^{2}}~\frac{M_{W}}{m_{d}^{2}-m_{s}^{2}}~\frac{m_{n}}{\pi}~\Lambda\nonumber\\
&=&\frac{38}{9 \pi^3}(G_{F} m_{N}^{2})^{2}\frac{m_{t}^{2}}{m_{s}^{2}}~\frac{m_{N}^{2}}{M_{W}^{2}}~\frac{\Lambda}
{m_{N}^{4}}~(\textrm{Im} V) \frac{e}{m_{n}}.
\end{eqnarray}

The reparametrization invariant combinations of the CKM matrix elements are also contained 
by the amount of CP violation in mixings and decays of the $K^{0}$ and $B^{0}$ systems \cite{mix}.
For instance, assuming no new physics contribution, the experimental bound on the $K^0-\overline{K^0}$ 
mixing parameter requires \cite{pakvasa}
\begin{equation}
\label{kk}
c_{1}s_{1}^{2}c_{2}s_{2}c_{3}s_{3} sin\delta_{CKM}\sim  10^{-5}.
\end{equation}
Using this parameterization, and supplying the masses of the quarks, we arrive at a numerical estimate for
the neutron EDM :
\begin{equation}
d_{n}\sim 10^{-32}~{\rm e-cm}.
\end{equation}
It is clear that this number is seven orders of magnitude below the experimental upper limit. It is plausible 
that the existing bounds will not change drastically unless there arises completely new high--precision experiments.
Therefore, this gap of seven orders of magnitude is probably going to be closed by the physics beyond the electroweak 
theory.
\section{Remarks}
In this chapter we have summarized the calculation of the neutron EDM when only the phase of the CKM matrix is 
present. The calculation involves several rough estimates for the hadron wavefunctions so there is always a
certain amount of uncertainty in the theoretical prediction. However, the numerical result is seven orders
of magnitude smaller than the present experimental upper bound, and this gap is unlikely to be closed by 
the theoretical uncertainties. Even if one (accidentally) forgets about the kaon system CP violation 
strength (\ref{kk}) the result is still 2--3 orders of magnitude below the experimental upper bound. It is
in this sense that the EDM in the electroweak theory is much smaller than the (long--living) upper bound. 

Apart from two--doublet models \cite{2hdmEDM}) and supersymmetry \cite{susyEDM}, in close 
similarity to what has been done in this section, one can discuss the neutron EDM in extensions of the 
SM with more than three generations, say four generations. In this case there are three phases in 
the corresponding CKM matrix with more reparametrization invariant combinations. However, there 
is still strong constraints from $K^{0}$ and $B^{0}$ systems \cite{mix4} so that it is 
unlikely that the neutron EDM will be levelled to the experimental upper bound \cite{edm4}.

\chapter*{Conclusion}
\addcontentsline{toc}{chapter}{Conclusion}

In this work we have presented a brief review of the neutron EDM calculation in the SM. In doing this we have 
analyzed the contributions of the strong and electroweak interactions separately. It is clear that the two 
force laws imply diversely different values for the neutron EDM: 
$$\left(d_{n}\right)_{strong}/\left(d_{n}\right)_{exp}\sim 10^{9}~,\ \ \ \mbox{and}~\ \ \
\left(d_{n}\right)_{electroweak}/\left(d_{n}\right)_{exp}\sim 10^{-7}~.$$
Each interaction here has its own source for violating the CP symmetry. These sources have nothing in common
concerning their nature and function in the theory. That of the strong interactions arises from the nontriviality of the
QCD vacuum concerning its topological and nonperturbative structure. However, the CP violation source of the 
electroweak theory is related to the number of families (for two families no source of CP violation, for four 
families there are two more sources), and the strength and hierarchy of the intergenerational mixings. 

The importance of these two contributions lies in the fact that they will always be present 
irrespective of what kind of extension of the SM is considered. For instance, in the supersymmetric models
there are new sources of CP violation \cite{cpphase} which generally exceed the experimental bounds by three
orders of magnitude \cite{susyEDM}. However, it is still meaningless to discuss the supersymmetric contribution alone, as the 
QCD contribution is there to exceed the bounds by several orders of magnitude unless a Peccei--Quinn type
scenario is exploited (such as \cite{gluino} or \cite{susyEDM}). 

The EDM of the neutron receives contributions from strong as well as electroweak interactions as mentioned above. 
This remains true in any extension of the SM model. However, being devoid of any color degrees of freedom,
the electron EDM receives no contribution from the QCD angle. Indeed, it is known that the experimental
bounds on the neutron and electron EDM's differ by an order of magnitude as could be taken into account 
by their mass ratios; $m_u/m_e\sim 10$. This expectation is confirmed by the predictions of supersymmetry \cite{susyEDM}
and two--doublet models \cite{2hdmEDM}. It is here that one observes the "excess" nature of the QCD contribution
which spoils the existing bounds by several orders of magnitude. For any underlying model, consistency 
among the EDM's and CP violation in meson systems is prerequisite for any meaningful prediction.

\section*{Acknowledgements}
\addcontentsline{toc}{chapter}{Acknowledgements}
{\sf The author greatfully acknowledges fruitful discussions with D. A. Demir. She also
thanks F. Hussain and G. Thompson for their kind help.}\\
{\sf It is a pleasure to thank M. A. Virasoro (Director, ICTP),  IAEA, and UNESCO for giving the author an oppurtunity
to be benefited from the Diploma Programme at ICTP.} 
 

\end{document}